\documentclass{article}
\usepackage{spconf,amsmath,graphicx}

\usepackage{lineno,hyperref}

\usepackage{epsfig,subfigure}
\usepackage{amsmath}
\usepackage{amssymb}
\usepackage{graphicx}
\usepackage{setspace}
\usepackage{multirow}
\usepackage[table]{xcolor}
\usepackage{amsmath,array,graphicx}
\modulolinenumbers[5]
\usepackage[german]{babel}
\usepackage{scalerel}

\usepackage{float}
\usepackage{stfloats}

\usepackage{textcomp}

\usepackage{graphicx}
\usepackage{booktabs}

\usepackage{array,tabularx}
\usepackage{gensymb}

\usepackage[
		     sort&compress]{natbib}
		     		     
\usepackage[table]{xcolor}
\usepackage{makecell}

\usepackage[symbol]{footmisc}

\usepackage{enumitem}

\makeatletter

\renewcommand\paragraph{\@startsection{paragraph}{4}{\z@}%
                                     {-3.25ex\@plus -1ex \@minus -.2ex}%
                                     {1.5ex \@plus .2ex}%
                                     {\normalfont\normalsize\bfseries}}
                                     
\usepackage{textgreek}

\DeclareMathOperator*{\argmax}{arg\,max}
\usepackage{siunitx}
\let\footnotemark

\begin{document}

\title{Uncertainty-driven ensembles of deep architectures for multiclass classification. Application to COVID-19 diagnosis in chest X-Ray images}

\def\@name{ \emph{Juan E. Arco$^{1,*}$\thanks{\textsuperscript{*} Corresponding author: jearco@ugr.es}},  \emph{Andr\'es Ortiz$^{2}$}, \emph{Javier Ram\'irez$^{1}$}, \emph{Francisco J. Mart\'inez-Murcia$^{2}$}, \\  \emph{Yu-Dong Zhang$^{3}$}, \emph{Juan M. G\'orriz$^{1}$}}


\address{\normalsize $^{1}$ Department of Signal Theory, Networking and Communications, Universidad de Granada\\ 
\normalsize $^{2}$ Department of Signal Theory, Networking and Communications, Universidad de Malaga \\
\normalsize $^{3}$ School of Informatics, University of Leicester, Leicester, LE1 7RH, Leicestershire, UK \\
}

\maketitle

\begin{abstract}
Respiratory diseases kill million of people each year. Diagnosis of these pathologies is a manual, time-consuming process that has inter and intra-observer variability, delaying diagnosis and treatment. The recent COVID-19 pandemic has demonstrated the need of developing systems to automatize the diagnosis of pneumonia, whilst Convolutional Neural Network (CNNs) have proved to be an excellent option for the automatic classification of medical images. However, given the need of providing a confidence classification in this context it is crucial to quantify the reliability of the model's predictions. In this work, we propose a multi-level ensemble classification system based on a Bayesian Deep Learning approach in order to maximize performance while quantifying the uncertainty of each classification decision. This tool combines the information extracted from different architectures by weighting their results according to the uncertainty of their predictions. Performance of the Bayesian network is evaluated in a real scenario where simultaneously differentiating between four different pathologies: control \textit{vs} bacterial pneumonia \textit{vs} viral pneumonia \textit{vs} COVID-19 pneumonia. A three-level decision tree is employed to divide the 4-class classification into three binary classifications, yielding an accuracy of 98.06\% and overcoming the results obtained by recent literature. The reduced preprocessing needed for obtaining this high performance, in addition to the information provided about the reliability of the predictions evidence the applicability of the system to be used as an aid for clinicians.

\end{abstract}


\section{Introduction}
\label{sec:intro}

Respiratory illness is the most common cause of death and disability in the world. According to the World Health Organization (WHO), tuberculosis kills 1.4 million people each year, whereas pneumonia is a leading cause of death among children under 5 years old \citep{who1}. Although the rate of pneumonia is decreasing worldwide \citep{who2}, an annual fatality rate of approximately 4 million is still observed. This disease is a form of acute respiratory infection that affects lungs, and based on the infectious pathogen, it can be bacterial, viral and fungal \citep{neu1}. Doctors can identify the presence of pneumonia from a wide range of medical imaging such as computed tomography (CT) \citep{ct}, chest X-ray (CXR) \citep{x-ray} or magnetic resonance imaging (MRI) \citep{mri}. The quality improvement in X-ray imaging   and its low cost has popularized the use of CXR as a diagnostic tool for pneumonia. However, this is not a straightforward task and success in pneumonia detection depends on many factors. One of the most important ones is that diagnosis is still largely dependent on the expertise of the radiologist \citep{chandra2020}. The pathology associated with pneumonia is often overlapping with other abnormal conditions of the lungs. Besides, the complex and vague anatomical structures in the lung fields can also affect the expert's opinion \citep{maduskar2016}. This leads to a manual, time-consuming process that has inter and intra-observer variability, which may delay diagnosis and treatment. The use of image processing methods along with machine learning algorithms directed to find disease-related patterns play a decisive role in the improvement of the diagnosis accuracy.

Previous works have employed machine learning (ML) algorithms for the automatic detection of a wide range of pathologies such as Parkinson's or Alzheimer's disease (\citealp{gorrizjm2020artificial,ZHANG2020149,castillo2018}), and most recently, pneumonia \citep{zhang2020_1,WANG2021208,chandra2020,elaziz2020}. In this direction, CAD (computer-aided diagnosis) systems can be an excellent tool for overcoming the weakness of current procedures for detecting pneumonia. In fact, they can assist radiologists by reducing their workload, serving as an B-reader in diagnosis and reducing the variability across doctors. Classification systems employed in CAD tools have the following general structure: i) delimitation of the regions of interest (ROI) to focus the analysis on them, ii) features extraction from these regions, iii) classification based on those features \citep{xu2006,jaeger2014,caixia2020}. Since pneumonia affects lungs, it seems obvious that the ROI must delimit the shape and boundaries of lungs \citep{van2001,hogeweg2015}. Several studies have provided different methods for lung segmentation \citep{akhila2017,candemir2014,munirah2015,guan2020,yang2018,vajda2018,donia2013}. \cite{munirah2015} proposed an unsupervised approach based on Gaussian derivatives filters and Fuzzy C-Means clustering. This method demonstrated not only good performance measures (accuracy of 0.9) but also robustness and speed. Regardless how features are computed they are then classified using a specific algorithm. Previous studies have successfully employed a wide range of classifiers for the detection of pulmonar diseases, such as DT (Decision Tree,\citealp{porcel2008,zhang2020}), NB (Na\"{i}ve Bayes, \citealp{chapman2011,ma2015}), or KNN (\textit{k}-Nearest Neighbors (KNN), \citealp{ajin2017,chen2015}). However, literature has shown that Support Vector Machine (SVM) \citep{yahyaoui2018,pan2018} usually outperforms the other algorithms \citep{uppaluri1999,chandra2020}.


Unlike classical methods based on the extraction of predefined features, deep neural networks build a specific feature space for the optimal class separation by means of a learning process. The emergence of these approaches has revolutionized the automatic classification of medical images. Recently, a number of studies have demonstrated the high flexibility and performance that this approach provides \citep{wang2017,varshni2019,kermany2018,mittal2020}. \cite{rajpurkar2017} proposed a 121-layer convolutional neural network (CNN) to identify pneumonia and localize the most indicative areas of this pathology. The algorithm provided a relatively low accuracy (76.8\%), but it was able to distinguish between 14 different pathologies. Other works have utilized transfer learning on the ImageNet dataset, yielding an accuracy of 82\%, 87\% and 92\% for Xception, VGG16 and VGG19 models, respectively \citep{abiyev2018}. It is clear that deep learning models can effectively identify the presence of a certain pathology. However, there are some scenarios where they take a decision (i.e. if a patient suffers from pneumonia or not) even though they do not know the answer since the classification outcome only relies on the most activated neuron of the output layer. \cite{kendall2017uncertainties} demonstrated the need of evaluating the uncertainty of a model's predictions in order to improve the decisions of the system. Bayesian deep learning models offer a practical solution for understanding the uncertainty of the decisions of a deep learning model \citep{ygal}. Specifically, they model a combination of aleatoric and epistemic uncertainty in order to increase loss robustness to noisy data, which usually leads to a boost in performance \citep{kendall2017uncertainties}. Most importantly, the additional information related to the reliability of the classification results makes this alternative quite interesting for being used in situations where the consequences of an error could be critical.

The recent COVID-19 pandemic has demonstrated the need of developing systems to automatize the diagnosis of pneumonia. It seems clear that a wrong diagnosis can have a dramatic effect in patient's health. In this work, we employ an ensemble classification system based on a Bayesian Deep Learning approach in order to maximize performance while quantifying the uncertainty of each classification decision. In particular, we combine seven CNN with the same structure, but differing in the kernel sizes of their convolutional layers. This allows the classification system to extract relevant features of different size and shape. The global classification is performed by combining the predictions of the different classifiers. The contribution of each individual classifier depends on the uncertainty of their predictions: the lower the uncertainty, the higher the weight, and vice versa. Performance of the Bayesian network is evaluated in a range of real scenarios of incremental difficulty: from the simplest one where trying to distinguish between control \textit{vs} pneumonia patients to a multiclass context where simultaneously differentiating between four different pathologies: control \textit{vs} bacterial pneumonia \textit{vs} viral pneumonia \textit{vs} COVID-19 pneumonia. The main contributions of our work can be summarized as follows:

\begin{itemize}
\item{A novel and accurate tool for the automatic diagnosis of pneumonia, in addition to the identification of the cause of the pathology (bacteria, virus, COVID-19).}
\item{The Bayesian nature of the Residual Network proposed in this work quantifies the reliability of the classification predictions.}
\item{The combination of networks with different kernel sizes allows the identification of pneumonia patterns regardless of their shape and extension.}
\item{Our approach employs the uncertainty of the predictions of each individual classifier to weigh their contribution to the ensemble global decision.}
\end{itemize}

\section{Material}
\label{sec:materials}

\subsection{Dataset}
\label{subsec:dataset}
\begin{figure*}
\centering
\includegraphics[width=0.8\textwidth]{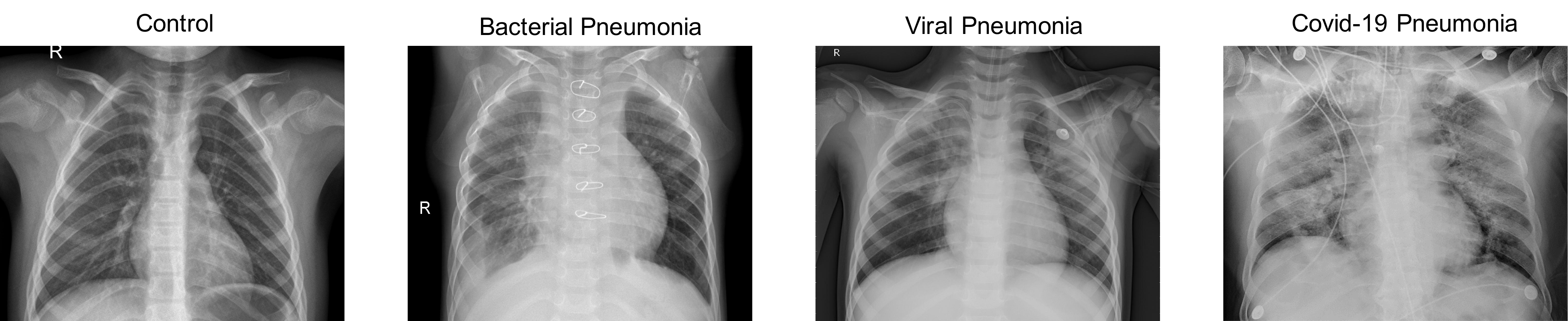}
\caption{From left to right, CXR image of a control, bacterial pneumonia, viral pneumonia and COVID-19 pneumonia. Note some clear artifacts in COVID-19 image.}
\label{fig:figurauno}
\end{figure*}

We have used the dataset available in \cite{kaggle} for controls and patients who suffered from a bacterial or a no-COVID19 pneumonia. According to the information described in \cite{kermany2018}, the CXR images were selected from retrospective cohorts of pediatric patients of one to five years old from Guangzhou Women and Children's Medical Center, Guangzhou. All CXR images were obtained as part of patient's routines clinical care. Institutional Review Board (IRB)/Ethics Committee approvals were obtained. The work was conducted in a manner compliant with the United States Health Insurance Portability and Accountability Act (HIPAA) and was adherent to the tenets of the Declaration of Helsinki. \cite{kermany2018} collected and labeled a total of 6374 CXR images from children, including 4273 characterized as depicting pneumonia and 1583 normal. From those patients diagnosed with pneumonia, 2786 were labeled as bacterial pneumonia, whereas 1487 were labeled as viral pneumonia. The dataset containing COVID-19 patients is available in \cite{kaggle1} and includes 576 CXR images from adults. Figure \ref{fig:figurauno} shows the CXR image from a control (CTL), and a patient suffering from a bacterial (BAC), a viral (VIR) and a COVID19 (CVD19) pneumonia.

\subsection{Image preprocessing}
\label{subsec:prepro}
When working with medical images, it is crucial to apply a preprocessing that improves the subsequent classification performance.This is especially important in CXR images, where low X-ray radiation and movement during image acquisition result in noisy and low-resolution images. However, this preprocessing must adapt images to the needs of the neural network. Due to computational and memory requirements, we downsampled the input images to obtain a final map of size 224x224. We also performed an intensity normalization procedure for each individual image based on standardization. Each image was transformed such the resulting distribution has a mean (\begin{math} \mu \end{math}) of 0 and a standard deviation (\begin{math} \sigma \end{math}) of 1, as follows:

\begin{equation}
I' = \frac{I-\mu}{\sigma}
\label{eq:clahe3}
\end{equation}

\noindent where \begin{math} I \end{math} is the original image and I' is the resulting one.

\section{Methods}
\label{sec:methods}

\subsection{Deep learning}
\label{subsec:deep}
The use of algorithms based on deep learning has revolutionized the analysis of medical images \citep{cnn1,cnn2,cnn3,cnn4}. From the ImageNet classification benchmark \citep{cnn5}, CNNs have been used more than any other pattern recognition algorithm in medical image classification. This architecture emerged as an attempt of replicating the behavior of neurons. Briefly, CNNs combine different steps based on convolution and pooling to allow the identification of different patterns and low and high-level features \citep{cnn6,cnn1}. The main component of a CNN is known as convolutional layer. This operator takes the tensor \begin{math} \mathbf{V}_{i-1} \end{math} containing the activation map of the previous layer \begin{math} i-1 \end{math}. Thus, the target layer (\begin{math} i \end{math}) learns a set of \begin{math} N \end{math} filters \begin{math} \mathbf{W}_i \end{math} with a bias term \begin{math} \mathbf{b}_i \end{math}, as follows:

\begin{equation}
\label{eq:cnn1}
\mathbf{V}_i = f_{a}(\mathbf{W}_i*\mathbf{V}_{i-1}+\mathbf{b}_i)
\end{equation}

\noindent where \begin{math} f_a(*) \end{math} is the activation function \citep{cnn1}. For a three-dimensional environment (\begin{math} \mathbf{V}_{i-1}\end{math}) of size \begin{math} H \times W \times D \times C \end{math} (height, width, depth and number of channels, respectively), \begin{math} \mathbf{W}_i \end{math} is of size \begin{math} P \times Q \times R \times S \times K \end{math} where \begin{math} K \end{math} is the number of filters. The \textit{k}th convolution term for the \textit{k}th filter is

\begin{equation}
\label{eq:cnn2}
\begin{split}
\mathbf{W}_{ik} * \mathbf{V}_{i-1} = \sum_{u=0}^{P-1} \sum_{v=0}^{Q-1} \sum_{w=0}^{R-1} [\mathbf{W}_{ik}(P-u, Q-v, R-w) \\ 
·\mathbf{V}_{i-1}(x+u,y+v,z+w)]
\end{split}
\end{equation}

Once convolution is performed, the activation of the filters in layer \begin{math} i \end{math} are stored and passed to the next layer \begin{math} i+1 \end{math}. It is of great importance to set properly the values for all the hyperparameters, striking a balance between performance and model complexity. One of these parameters is the number of filters: the higher this number is, the more patterns the model is able to learn. There is no consensus in literature about the ideal number of filters, probably because different problems need CNNs with different configurations, but numbers that are a power of 2 are usually taken.

\subsection{Bayesian Deep learning}
\label{subsec:bay_deep}
Despite the high performance that Deep Learning models have demonstrated, recent works have claimed the need of computing the uncertainty of a model, a measure that allows to identify situations where the classifier does not know the answer. To do so, it would be necessary to estimate the level of uncertainty of a prediction in order to reject it in case its value was too high. Bayesian deep learning offers a framework for understanding uncertainty with deep learning models \citep{bdl2016}. There are two main types of uncertainty that can be estimated in Bayesian modeling: epistemic and aleatoric \citep{KIUREGHIAN2009105,ygal}. Epistemic uncertainty is inherent to the model, which means that it can be reduced by increasing the data processed by the model. Estimating the epistemic uncertainty requires to model distributions over the different parameters of the model. This allows to optimize the network according to the average of all possible weights. 

Let \begin{math} \mathbf{x} \end{math} be a feature vector and \begin{math} \mathbf{W} \end{math} the weights of a Bayesian Neural Network (BNN). Considering the output of the network as \begin{math} \mathbf{f}^{\mathbf{W}}(\mathbf{x}) \end{math}, the model likelihood can be defined as \begin{math}p(\mathbf{y}|\mathbf{f}^{\mathbf{W}}(\mathbf{x})) \end{math}. For a given dataset \begin{math} \mathbf{X} = \{\mathbf{x}_{1},\cdots, \mathbf{x}_{N}\} \end{math}, \begin{math} \mathbf{Y} = \{\mathbf{y}_{1},\cdots, \mathbf{y}_{N} \} \end{math}, the Bayesian inference computes the posterior probability over the weights \begin{math} p(\mathbf{W}|\mathbf{X},\mathbf{Y}) \end{math}.

\cite{cipolla_unc} demonstrated that applying dropout before every weight layer in a neural network is mathematically equivalent to an approximation to the probabilistic deep Gaussian process \citep{damianou2013}. Briefly, they showed that the dropout objective minimizes the Kullback-Leibler divergence between an approximate distribution and the posterior one of a deep Gaussian process. A popular technique relies on the use of Monte Carlo dropout sampling to place a Bernoulli distribution over the network's weights. Dropout is widely used as a regularization procedure during training \citep{cnn8}. However, when applied during the testing phase, this method allows to obtain a distribution for the output predictions \citep{jospin2020,dropout2017}. The statistics of this distribution reflect the model's epistemic uncertainty. 

%
%

\begin{figure*}
\centering
\includegraphics[width=0.7\textwidth]{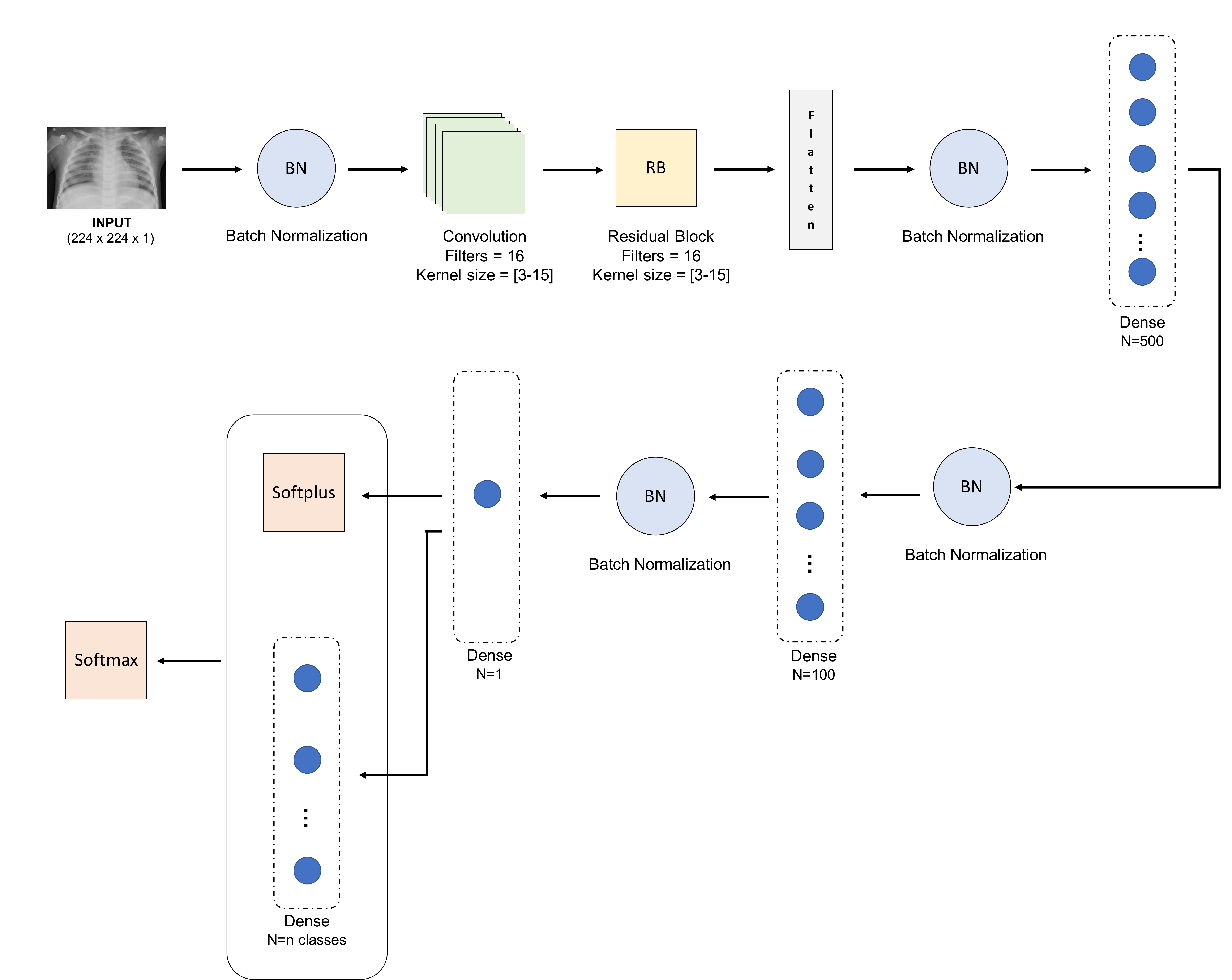}
\caption{Diagram of the bayesian framework of each individual network within the ensemble.}
\label{fig:bayesian_network}
\end{figure*}

Aleatoric uncertainty is usually referred as the uncertainty inherent to the data, and can be divided into two sub-categories: i) homoscedatic uncertainty, which remains stable for every input of the model; and ii) heteroscedastic, which assumes that noise varies for the different inputs of the model \citep{nix1994,lequoc2005}. Heteroscedastic uncertainty can be modeled by modifying the loss function used by the neural network. Since this uncertainty is a function of the input data, employing a deterministic mapping from inputs to model outputs can allow the estimation of the uncertainty. For a typical Euclidean loss \begin{math} L = ||y-\hat y||^2 \end{math}, the Bayesian version will be given by \begin{math} L = \frac{||y-\hat y||^2} {2\sigma ^2} + \frac{1}{2} \log \sigma ^2 \end{math}. In the latter one, the model predicts both \begin{math} \hat{y} \end{math} and variance \begin{math} \sigma^2 \end{math}, so that if model prediction is not good, the residual term will be attenuated by increasing \begin{math} \sigma^2 \end{math}. Therefore, the term \begin{math} \log \sigma^2\end{math} prevents uncertainty growing until infinite, leading to a learned loss attenuation. The process for homoscedastic uncertainty is essentially the same, but considering the uncertainty like a free parameter instead of a model output.

%

In this work, we have employed a Bayesian version of the ResNet-18 CNN \citep{he_zhang}. The output layer contained 2 neurons with softmax activation. Besides, dropout was used to prevent overfitting, and Batch Normalization for convergence. The Bayesian nature of this net is obtained by replacing the deterministic weights along the network by a distribution over these parameters.This means that instead of optimising the network weights, an average of all possible weights was computed. As a result, the loss function depends on two factors: the softmax values (as in the non-Bayesian modality) and the Bayesian categorical cross entropy, which is based on the input variance (see \cite{kendall2017uncertainties} for more details). Figure \ref{fig:bayesian_network} summarizes the architecture of the Bayesian network.

\subsection{Multi-level Ensemble Classification}
\label{subsec:ensemble}

Patterns associated with each type of pneumonia are similar among different subjects. However, there are some factors like the virulence of the disease and the presence of other pulmonary findings that can affect the identification of the patterns associated with the different pathologies. One crucial aspect is to select an optimal kernel size for the convolutional operators of the neural network that can properly extract the relevant information. Moreover, this is even more important when images used to train the network come from different sources, and when they can have different sizes and aleatory artifacts. To overcome this issue, we employed seven neural networks, each one of them with a different kernel size value in the range \begin{math} [3-15] \end{math} with increments of two. This means that the kernel size assigned to the first network was 3, 5 for the second network and so on, until a size of 15 for the seventh CNN. The number of neural networks and their kernel sizes were selected in order to strike a balance between performance and computational cost. Finally, each individual classifier was then combined into a global one following an ensemble classification procedure. 

\begin{figure*}
\centering
\includegraphics[width=0.7\textwidth]{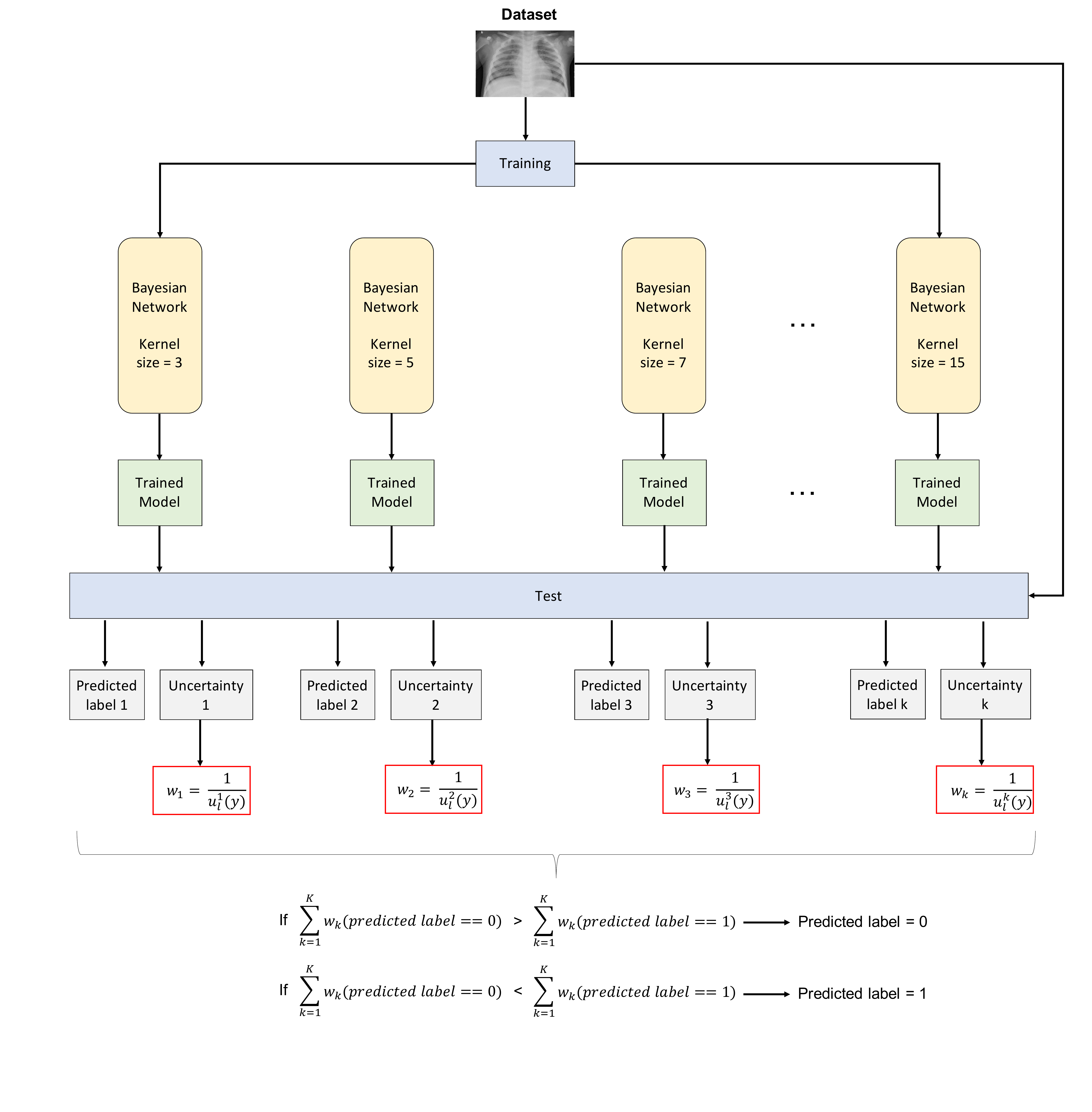}
\caption{Schema of the ensemble architecture proposed in this work based on the uncertainty in the prediction of each individual classifier.}
\label{fig:ensemble}
\end{figure*}

Previous studies have employed majority voting to fuse the output of the base classifiers \citep{chandra2021,zhou2020}. Given the Bayesian nature of the networks employed in this work, we computed the weights of each classifier as a function of the uncertainty given by each one of them for each test image (see Figure \ref{fig:ensemble}). If the uncertainty of a classifier in a specific prediction was high, it would have a low contribution to the final ensemble, and vice versa \citep{ensemble2012}. Defining \begin{math} u_{l}^{k}(\mathbf{y}) \end{math} as the uncertainty of the test sample \begin{math} \mathbf{y} \end{math} obtained from the \textit{k}-th classifier corresponding to the \textit{l}-th class, the empirical average of the \textit{l}-th weights (inverse of uncertainties) over the \textit{K} classifiers can be calculated as follows:

\begin{equation}
\label{eq:ensemble1}
E_{l}(\mathbf{y}) = \frac{\sum_{k=1}^{K}{\frac{1}{u_{l}^{k}(\mathbf{y})}}}{K}
\end{equation}

The class label of the test sample \begin{math} \mathbf{y} \end{math} is then assigned to the class with the maximum average weight as:

\begin{equation}
\label{eq:ensemble2}
Label(\mathbf{y}) = \argmax_{l} \hspace{0.5mm} E_{l}(\mathbf{y})
\end{equation}

Detecting the presence of pneumonia when comparing to healthy subjects is an interesting initial step in the development of a CAD system. However, it is much more useful to identify exactly the type of pneumonia patients suffer from. As described in Section \ref{subsec:dataset}, the database used in this work contains CXR images of healthy subjects (controls) and images from three types of pneumonia: bacterial, viral and COVID-19. In order to perform the multiclass classification, we employed a decision tree based on the One-versus-all (OVA) approach \citep{multiclass1,multiclass2,multiclass3}. This alternative divides a multiclass problem into a number of binary sub-problems. In each one of them, one of the classes is considered as the positive class, whereas the other classes are the negative class. Following this framework, we used a decision tree with three levels in order to distinguish between the different pathologies. In each level, an ensemble of different kernel sizes was employed. This led to a two-level ensemble classification: one ensemble for the combination of different kernels, and another one for combining binary classifiers to perform multiclass classification. The decision tree relies on a process that can be summarized as follows:

\begin{itemize}
\item{First level: classification between normal \textit{vs} pneumonia. The second class contains subjects diagnosed from the three different types of pneumonia (bacterial, viral, and COVID-19)}. 
\item{Second level: classification between bacterial \textit{vs} viral pneumonia. The second class corresponds to images from subjects with pneumonia due to different viruses (no-COVID-19 or COVID-19)}.
\item{Third level: classification between no-COVID-19 \textit{vs} COVID-19.}
\end{itemize}

\begin{figure*}
\centering
\includegraphics[width=0.7\textwidth]{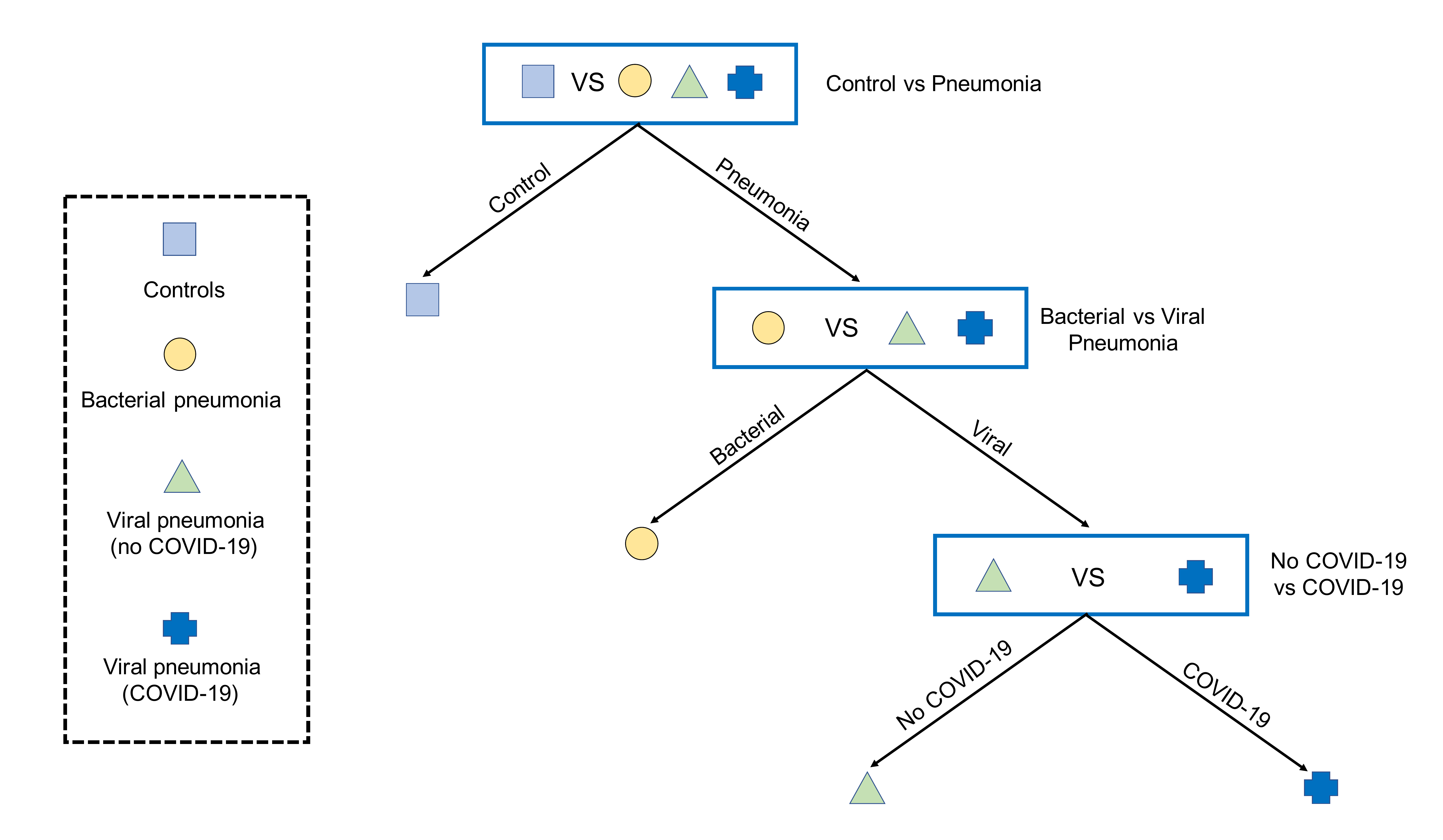}
\caption{Schematic representation of the decision tree employed for the multiclass classification.}
\label{fig:decision_tree}
\end{figure*}

Figure \ref{fig:decision_tree} depicts a visual representation of how the decision tree works. Images that are labelled as pneumonia in the first level are passed to the second one. Similarly, images labelled as viral pneumonia continue to the third level in order to identify whether the virus that produced the pneumonia was COVID-19 or not. It is worth mentioning that the binary classifier employed in each level has the same ensemble structure that the one explained in Section \ref{subsec:ensemble}.

\subsection{Performance evaluation}
\label{subsec:performance}
For all experiments, a 5-fold stratified cross-validation scheme was used to estime the generalization ability of our method \citep{Kohavi95}. The performance of the classification frameworks was evaluated in terms of different parameters from the confusion matrix, which can be computed as follows:

\begin{align}
\label{eq:metrics}
\begin{split}
Acc = \frac {T_{P}+T_{N}}{T_{P}+T_{N} + F_{P}+F_{N}}  \hspace{0.3cm} Sens = \frac {T_{P}}{T_{P}+F_{N}}  
\\
Spec = \frac {T_{N}}{T_{N}+F_{P}} \hspace{0.5cm} AUC = \frac{1}{2} \Big( \frac{TP}{P} + \frac{TN}{N}\Big) 
\\
Prec = \frac {T_{P}}{T_{P}+F_{P}}  \hspace{0.5cm} F1-score = \frac{2 \times Prec \times Sens}{Prec + Sens}
\end{split}
\end{align}

\noindent where \begin{math}  T_{P} \end{math} is the number of pneumonia patients correctly classified (true positives), \begin{math} T_{N} \end{math} is the number of control patients correctly classified (true negatives), \begin{math} F_{P} \end{math} is the number of control subjects classified as pneumonia (false positives) and \begin{math} F_{N} \end{math} is the number of pneumonia patients classified as controls (false negatives). We also employed the area under the curve ROC (AUC) as an additional measure of the classification performance \citep{auc1,auc2}. Since classes were unbalanced (e.g. the number of pneumonia patients was higher than controls), we incorporated the weights of the classes into the cost function in order to the majority class does not contribute more than the minority one.

Given the ensemble nature of the system proposed in this work, we employed a kappa-uncertainty diagram to evaluate the level of agreement of the different classifier outputs while correcting for chance \citep{rodriguez2006,wang2019}. This measure is based on Cohen's kappa coefficient \citep{cohen1960}, which is widely accepted as the de facto standard for measurement of interannotator agreement \citep{kappa_statistic2004}. Specifically, the kappa statistic compares an observed accuracy with an accuracy obtained by chance, providing a measure of how closely instances classified by a classifier match the ground truth. Mathematically, Cohen's kappa can be defined as:

\begin{equation}
\label{eq:cohen}
k = \frac{p_A-p_E}{1-p_E}
\end{equation}

\noindent where \begin{math} p_A \end{math} is the observed relative agreeement between two annotators, and \begin{math} p_E\end{math} is the probability of agreement by chance. Although acceptable kappa statistic values vary on the context, the closer to 1, the better the classification. Section \ref{sec:results} summarizes the kappa scores obtained by different members of the ensemble classifier, as well as revealing the relationship between the uncertainty of Bayesian networks and kappa values.

As explained in Section \ref{subsec:ensemble}, a decision tree was employed for multiclass classification. In order to build the kappa-uncertainty diagram explained above, a combination of the uncertainties of the different levels of the tree has to be computed. To do so, we employed a method known as summation in quadrature \citep{unc_combination}, described as follows:

\begin{equation}
\label{eq:unc_combined}
u_{c}(y) = \sqrt{\sum_{i=1}^n [c_i u(x_i) ]^2}
\end{equation}

\noindent where \begin{math}u_{c}(y) \end{math} is the combined uncertainty, \begin{math} c_i \end{math} is the sensitivity coefficient and \begin{math} u(x_i)\end{math} is the standard uncertainty.

\section{Evaluation}
\label{sec:eval}

\begin{table*}[ht]
\caption{Performance of the ensemble classification approach proposed in this work in the different contexts evaluated.}
\label{table:results1}
\begin{tabular*}{\textwidth}{@{\extracolsep{\fill}}cccccccc|}
 \hline
 Experiment & Acc (\%) & Sens (\%) & Spec (\%) & Prec (\%) & AUC (\%) & F1-score (\%) \\
 \hline
CTRL vs PNEU  & 97.27 \textpm 3.37 & 96.41 \textpm 4.47 & 99.94 \textpm 0.13 & 99.98 \textpm 0.04  & 98.17 \textpm 2.22 & 98.11 \textpm 2.38\\
BAC vs VIR & 98.43 \textpm 0.95  & 98.16 \textpm 1.17 & 98.79 \textpm 0.73 & 99.09 \textpm 0.56 & 98.48 \textpm 0.92 & 98.62 \textpm 0.84\\
COVID-19 vs NO COVID-19  & 99.69 \textpm 0.56  & 99.83 \textpm 0.35 & 99.6 \textpm 0.8 & 99 \textpm 1.98 & 99.71 \textpm 0.4 & 99.4 \textpm 0.99\\
Multiclass & 98.06 \textpm 1.63 & 97.24 \textpm 2.67 & 99.38 \textpm 0.33 & 99.6 \textpm 0.21 & 98.31 \textpm 1.33 & 98.39 \textpm 1.38 \\
 \hline
\end{tabular*}
\end{table*}

\subsection{Experimental setup} 
\label{subsec:setup}
In this work we propose a method to extract the relevant information from CXR images that allows the identification of pneumonia. To do so, we define two experiments:

\begin{itemize}[leftmargin=*]
\item{\textbf{Experiment 1: Binary Classification} between different groups under three scenarios: \textbf{CTL \textit{vs} PNEU}}, which includes all images labelled as CTL and PNEU; \textbf{BAC \textit{vs} VIR}, which divides the images from people diagnosed from pneumonia regarding the cause of the disease is a bacteria or a virus; \textbf{NO-COVID-19 \textit{vs} COVID-19} for viral pneumonia. The aim is to identify whether the virus that produced pneumonia was COVID-19 or not. The whole Bayesian CNN was trained using the Adam optimization algorithm \citep{adam}, with learning rate 0.001, \begin{math} \phi = 0.9 \end{math} and a decay of 0.001). The number of epochs employed for training the system was 15, 20 and 25, for the CTL \textit{vs} PNEU, BAC \textit{vs} VIR and NO-COVID-19 \textit{vs} COVID-19 scenarios, respectively. We used the Keras library over Tensorflow with some custom modules.

\item{\textbf{Experiment 2: Multiclass Classification} by using a decision tree in order to distinguish between the four different pathologies contained in the database. A binary classification is employed in each of the three levels of the tree. The first level corresponds to the CTL \textit{vs} PNEU classification, the second one contains the BAC \textit{vs} VIR comparison, whereas in the third level, the distinction between NO-COVID-19 \textit{vs} COVID-19 is performed. These binary classifiers employ the same framework and configuration as in Experiment 1.}
\end{itemize}

\section{Results}
\label{sec:results}

\begin{figure*}
\centering
\includegraphics[width=0.68\textwidth]{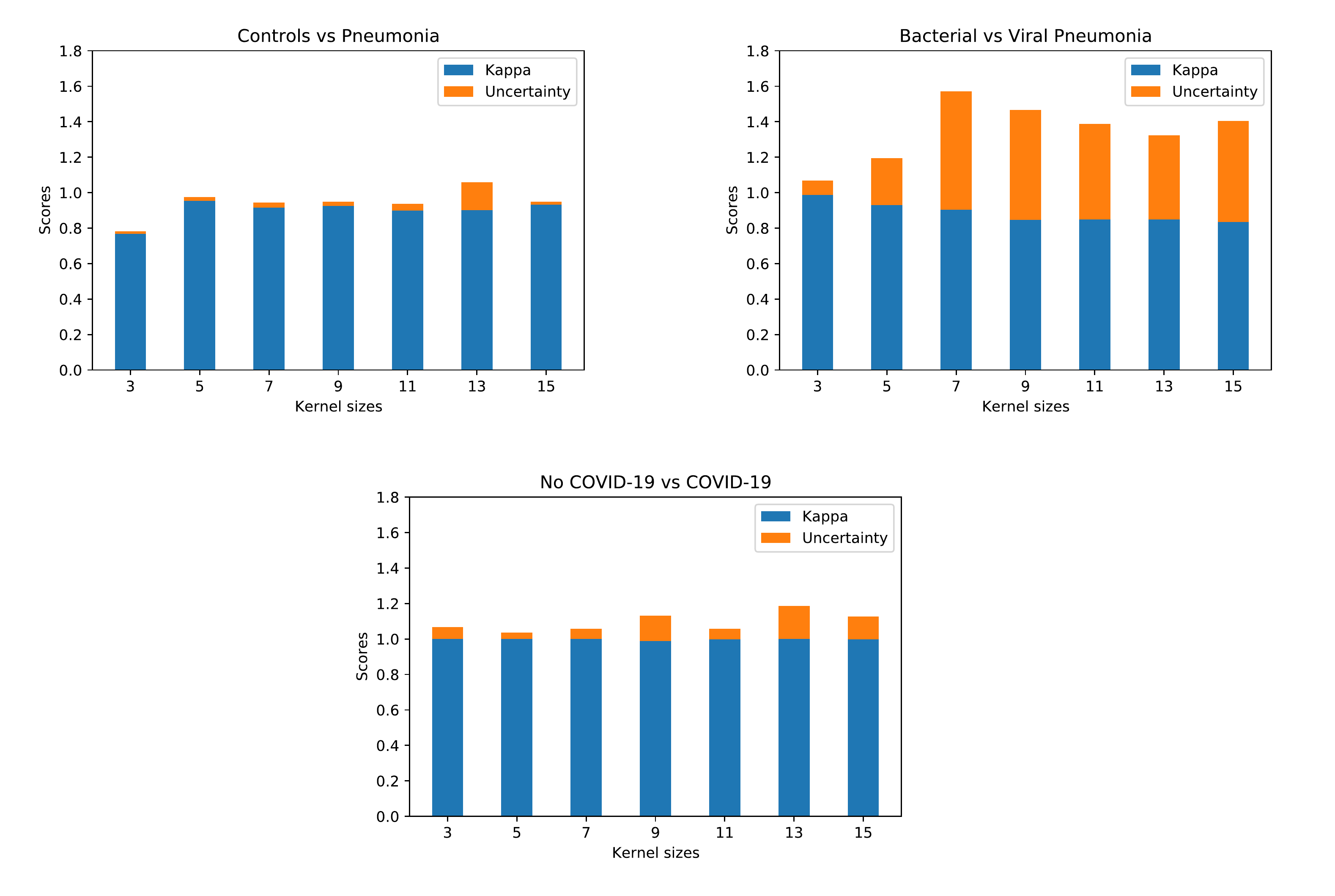}
\caption{Performance associated with the different kernel sizes for the three classification contexts under study. Scores evaluated were kappa and uncertainty.}
\label{fig:kernel_size}
\end{figure*}

We first explore how performance varies for the different kernel sizes of the individual classifiers for all the binary classifications performed (see Figure \ref{fig:kernel_size}). We can see that kappa score slightly varies when increasing the kernel size in the three classification contexts. With reference to uncertainty, only in the BAC \textit{vs} VIR scenario uncertainty values drastically change for different kernel sizes. Therefore, there is not a tendency that let us assure that there is a relationship between these two variables. It is important to note the high levels of uncertainty in this classification context when comparing to the first and the third one, which manifests the extreme difficulty of this specific classification. It is not surprising that differentiating between a control and a patient who suffer from pneumonia is a considerably easier task. However, these findings point out that there is a larger difference in the spatial patterns associated with COVID-19 \textit{vs} no-COVID-19 than in the one between bacterial \textit{vs} viral. This can be explained by the severity of the pulmonary affection that COVID-19 usually causes, whereas pneumonia derived from another viruses can show a more heterogeneous severity. 

\begin{figure*}
\centering
\includegraphics[width=0.42\textwidth]{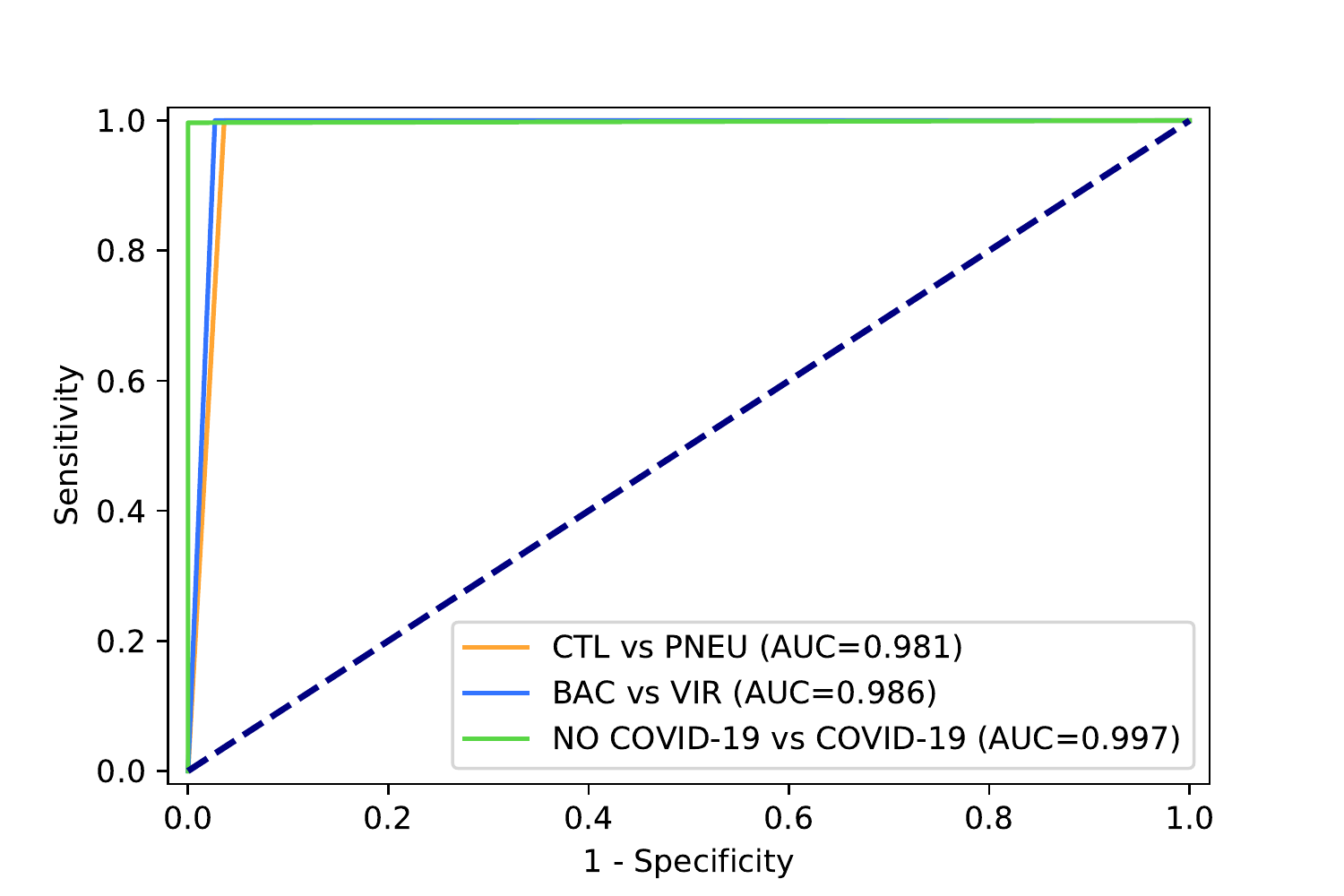}
\caption{ROC curves obtained by the classifiers of each level of the decision tree.}
\label{fig:roc_curves}
\end{figure*}

\begin{figure*}
\centering
\includegraphics[width=0.5\textwidth]{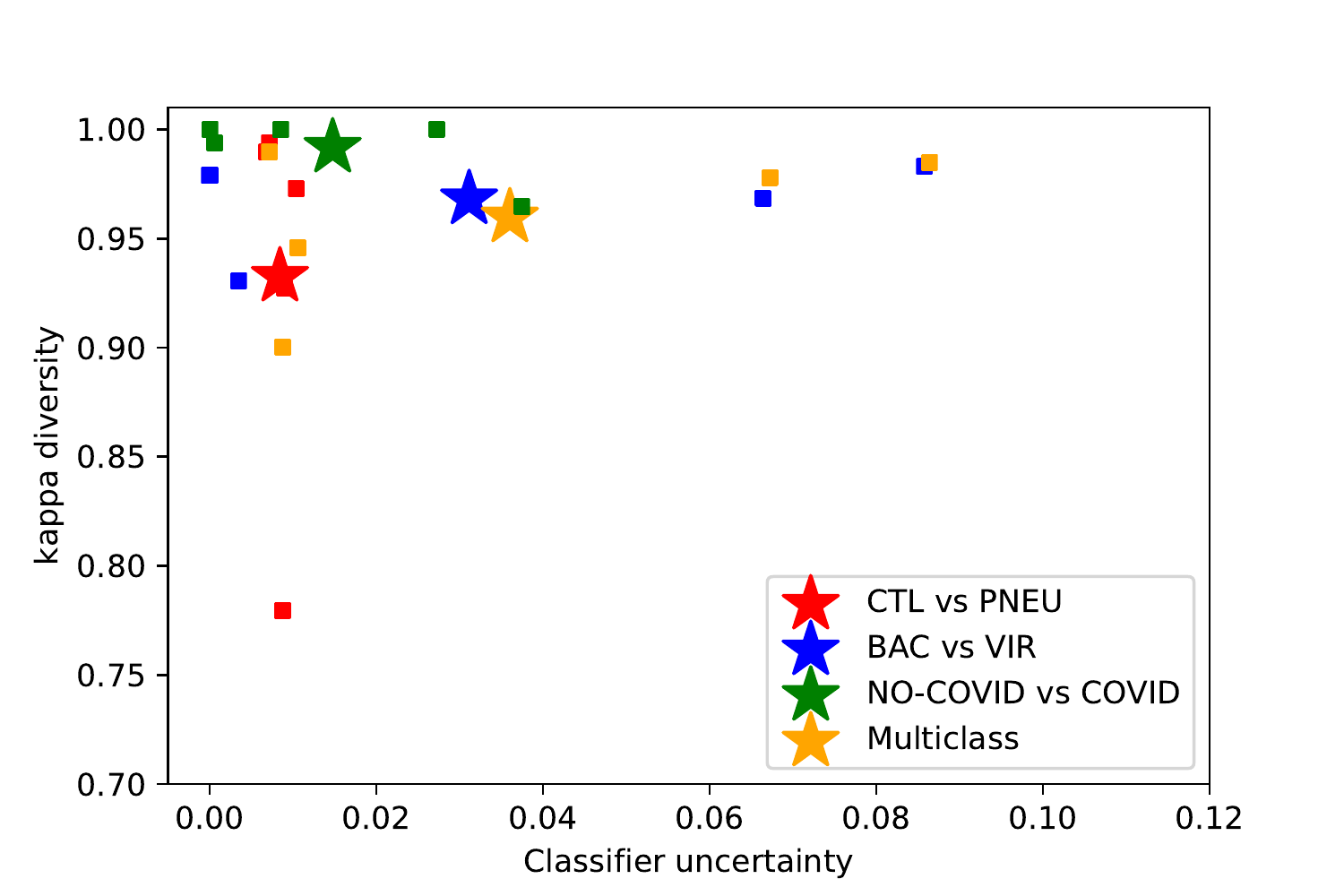}
\caption{Diversity-uncertainty diagrams of the different levels of the multiclass classifier. The x-axis represents the combined uncertainty of each individual classifier and the resulting mutlticlass. The y-axis represents diversity of the classifiers evaluated by the kappa measure. Each dot represents the kappa-uncertainty score obtained by a classifier in one fold, whereas large stars represent the centroid of the resulting distribution.}
\label{fig:kappa_unc}
\end{figure*}

We observe that the discrimination ability of the system is very high for the three binary classifications regardless of the kernel size employed . Results in terms of different performance measures are shown in  Table \ref{table:results1}, whereas Figure \ref{fig:roc_curves} depicts the ROC curves for the different classifiers. Large values are obtained, as expected, in the CTL \textit{vs} PNEU context. However, these results confirm that our system can also separate patients with the same diagnosis (pneumonia) but with a different cause (bacteria, virus, COVID-19). We also use the kappa-uncertainty diagram to evaluate the level of agreement between the classifier outputs. Figure \ref{fig:kappa_unc} shows these diagrams for the three binary classifiers and the multiclass derived from the decision tree, represented by a different colour. The cloud points represent the kappa score-uncertainty obtained in each fold of the cross-validation scheme, whereas large stars represent the centroid of the resulting distribution. From this figure, we can see that there is not a great difference between individual classifiers, in consonance with results derived from ROC curves. 

It is interesting how this figure reveals that the combination of classifiers with a certain performance (high kappa score and low uncertainty) leads to an ensemble classifier with these features. However, uncertainty is higher in the multiclass classifier for a similar kappa score compared to individual ones.This means that, although classification performance of the decision tree is high, the uncertainty of the resulting prediction is also higher than in binary classification. This evidences the extreme utility of this kind of diagrams in Bayesian deep learning and in contexts when reliability of predictions is of core interest. According to Table \ref{table:results1}, the multiclass classifier has a superior performance than the CTRL \textit{vs} PNEU in most of metrics evaluated. However, the uncertainty of the predictions is also higher (centroid of the multiclass is farther to the right than the CTRL \textit{vs} PNEU centroid). Further discussion regarding the results obtained and their clinical implications are provided in Section \ref{sec:discussion}.

\section{Discussion}
\label{sec:discussion}

In this study, we proposed a classification method for the detection of different types of pneumonia from CXR images. This approach relies on the use of a Bayesian version of a Residual Network (ResNet), which allows the optimization of the network according to the uncertainty of its predictions. We employed different networks modifying their kernel sizes and combined them within an ensemble classifier so that the contribution of each individual network depends on the uncertainty of its predictions. We evaluated the performance of this approach in different classification scenarios. In the first context, the two classes generated relatively big differences in the observed pattern (pneumonia \textit{vs} control), whereas in the second (bacterial \textit{vs} viral pneumonia) and in the third one (COVID-19 \textit{vs} no-COVID-19) these differences were extremely small. Besides, the performance of a multiclass classifier was also evaluated in order to check if this method could simultaneously differentiate between the different pathologies.


The high performance shown by the proposed method in all scenarios provides us with a new tool to detect the presence of pneumonia in CXR images, in addition to distinguish whether the source of the pathology is viral or bacterial, and if the virus is COVID-19 or not. The features extracted by convolutional blocks of different kernel sizes contained relevant information that enhanced the separability between the different classes. The combination of convolutional blocks of different kernel sizes is especially interesting in this context where the database contains images of people from a wide range of age. Pulmonary affections caused by the different pathologies evaluated in this work mainly depend on the severity of the disease. However, the shape and size of these manifestations also depend on the shape and size of lungs. The ensemble method proposed in this work allows the identification of patterns associated with pneumonia without focusing on a specific size for the informative regions. 

Another crucial aspect of the method proposed in this work is its Bayesian nature. The aim of CAD systems regardless of the application context is to maximize the classification performance, in terms of accuracy, AUC, etc. However, in most scenarios it is also important to know the reliability of the prediction itself. Neural networks are prone to overfitting, which means that taking decisions based only on the prediction can be counterproductive. In an extreme case, it is possible that the classifier does not know the class a test image belongs to, but it always has to assign a label, even though the output probability is near to chance level. This is particularly problematic when developing a tool for the diagnosis of a disease. Doctors need to know not only the global accuracy obtained during the training and test of the model, but how reliable is the prediction of new individual samples. This problem is addressed with the inclusion of Bayesian elements in neural networks. However, our findings reveal that this is not the only advantage that this approach provides. We have demonstrated the high performance of ensemble classification, even in situations where differences between the pulmonary patterns of the different pathologies are extremely small. The novelty of our approach relies on the way the contribution of each individual classifier to the global decision is computed. Weights are usually derived from the accuracy of each individual classifier. However, results can be biased if part of the predictions are obtained by chance, i.e. when the output probabilities of the different classes are almost equal. We overcome this problem by weighting the contribution of each classifier according to the uncertainty of its predictions.

It is worth remembering that part of the database (normal, bacterial and viral (no-COVID) pneumonia patients) contains pediatric chest radiographies, whereas COVID-19 images correspond to adults. Detecting pneumonia from pediatric chest radiographies is more challenging than in adults for several reasons. First, the dose of X-ray radiation is considerably lower than in adults, which results in a reduced image resolution and a higher overlapping between the different anatomical parts. Second, lungs appearance changes dramatically along the pediatric development stages, both in size and shape (more similar to a triangle in infants). The dataset employed in this work contains CXR images of children of a wide range of ages, increasing variability and complexity of the classification process. Finally, CXRs are noisier than in adults because of movement, legs positioning or when they are being hold by adult hands. For this reason, it is worth highlighting the high performance obtained in this work, improving the results obtained in previous works even when applied to detect pneumonia in children \citep{rajaraman2018,liang2020,measurement2020}, adults \citep{pneu_adults1,pneu_adults2,pneu_adults3}, and also when tried to identify the presence of COVID-19 \citep{covid1,covid2,covid_multiclass,covid3}. 

We have developed a tool that is able to distinguish between patterns associated with different pathologies, but it is worth highlighting the high performance obtained in the multiclass classification. In this case, the accuracy and the AUC obtained were 98.06\% and 98.31\%, respectively, which is considerably higher than the results provided by similar techniques in previous studies \citep{zhang2020_1,wang2020_ct,zhou2020,hemdan2020covidxnet,apostolopoulos2020}. There are two main relevant aspects derived from these excellent results to be mentioned. First, the only preprocessing applied to the data was the rescaling of the images to a lower resolution in order to reduce the computational burden of the classification pipeline. We did not perform other complex processes such as lung segmentation, but the RAW rescaled images were used as the inputs of the classification system. Thus, it is remarkable the high performance obtained by the method proposed in this extreme situation. Second, results obtained in the multiclass scenario allow the application of the tool proposed in this work in a more real context. The multiclass scenario is more similar to a real context than binary classifications, where the simplest case only distinguishes the presence (or not) of pneumonia. Results obtained by the multiclass classifier reveal the usefulness of this kind of techniques.

\section{Conclusion}
\label{sec:conclusion}
Respiratory illness is leading cause of death and disability in the world. The annual fatality rate of pneumonia is approximately 4 million people, whereas is the leading cause of death among children under 5 years old. The pathology associated with pneumonia is often overlapping with other abnormal conditions of the lung, leading to a time-consuming process that may delay diagnosis and treatment. In this paper we proposed an uncertainty-driven ensemble of deep neural networks to identify patterns associated with different types of pneumonia. This tool combined the information extracted from different architectures according to the uncertainty of their predictions, instead of using the accuracy of individual classifiers as most studies usually do. The information provided about the reliability of the predictions, in addition to the large performance obtained (accuracy of 98.06\% when distinguished between four pathologies) evidences the applicability of the system to be used as an aid for clinicians. The combination of CNNs of different kernel sizes allows the identification of pneumonia patterns regardless of their size and shape. Moreover, the reduced preprocessing needed for obtaining these results guarantees a limited computational cost. Our results pave the way for the application of Bayesian deep neural networks to other image modalities such as CT, which offers much more resolution than XR images and can provide key information for the detection of pneumonia.

\section*{Acknowledgments}\label{sec:Acknowledgments}
This work was partly supported by the MINECO/ FEDER under the PGC2018-098813-B-C32, RTI2018-098913-B100, CV20-45250 and A-TIC-080-UGR18 projects.

%
%
%
%

\vfill
\pagebreak

\bibliographystyle{elsarticle/elsarticle-harv}

\bibliography{biblio}

\end{document}